\documentclass[11pt]{article}
\usepackage{amsmath,amsfonts,amsbsy,amssymb}
\usepackage{bm}
\usepackage{cite}
\usepackage{graphicx,epsfig}
\usepackage{geometry}
\makeatletter
\renewenvironment{thebibliography}[1]
{\section*{\refname\@mkboth{\refname}{\refname}}%
  \list{\@biblabel{\@arabic\c@enumiv}}%
       {\settowidth\labelwidth{\@biblabel{#1}}%
        \leftmargin\labelwidth
        \advance\leftmargin\labelsep
 \setlength\baselineskip{11pt}%
        \@openbib@code
        \usecounter{enumiv}%
        \let\p@enumiv\@empty
        \renewcommand\theenumiv{\@arabic\c@enumiv}}%
  \sloppy
  \clubpenalty4000
  \@clubpenalty\clubpenalty
  \widowpenalty4000%
  \sfcode`\.\@m}
 {\def\@noitemerr
 {\@latex@warning{Empty `thebibliography' environment}}%
\endlist}
\makeatother
\def\vec#1{\boldsymbol{#1}}
\begin{document}
\centerline{{\sl Genshikaku Kenkyu Suppl.} No. 000 (2012)}
\begin{center} 
\vskip 2mm
{\Large\bf

The Role of Flavor in Multiquark Spectroscopy 
\hspace{-1mm}\footnote{Presented at the International Workshop on Strangeness 
Nuclear Physics (SNP12), August 27 - 29, 2012, \\
\hspace*{5mm} Neyagawa, Osaka, Japan.}
}\vspace{5mm}

{
Jean-Marc Richard$^a$
}\bigskip

{\small
$^a$Universit\'e de Lyon, Institut de Physique Nucl\'eaire, IN2P3-CNRS--UCBL, \\
4, rue Enrico Fermi, 69622 Villeurbanne, France
}
\end{center}
\vspace{3mm}

\noindent
{\small \textbf{Abstract}:\quad
In several mechanisms proposed for multiquark binding, the flavor content of the quark configuration plays a
very important role. We shall review some past or recent results on multiquarks dealing with the chromomagnetic interaction and with the spin-independent potential among quarks.
}%


\section{Introduction}

There is a persisting activity in multiquark spectroscopy, in spite of the highs and lows of the experimental search.
This physics is now studied in the framework of QCD sum rules and lattice QCD, but most contributions have been worked out within simple constituent models. The difficulty there is twofold. First, potentials that are successful for mesons and baryons should be extrapolated towards the multiquark sector. Second, the four-, five-, or six-body problem has to be solved accurately to decide whether the ground state is a compact multiquark, or just the threshold made of two separated hadrons.

In this contribution, we shall come back on the history of the $H$ dibaryon, the heavy pentaquark $P$ and the prediction of $T_{QQ}$ tetraquarks with two heavy quarks. Unfortunately, it will not be possible to discuss in detail the recent states $X$, $Y$, $Z$, etc. with hidden charm and beauty. We refer to recent reviews such as \cite{Swanson:2006st,Brambilla:2010cs,Nielsen:2009uh}. For other surveys of multiquark physics, see, for instance, \cite{Jaffe:2004ph,Lipkin:2007cg}. For an introduction to the quark model, see, e.g., \cite{Richard:2012xw}.

\section{Chromomagnetism}

In QED, the spin-dependent corrections to the one-photon exchange gives a very successful account for the fine and hyperfine interactions in atoms, in particular the celebrated 21\,cm line of hydrogen. 
It was early recognized that, similarly, the spin-dependent terms associated with one-gluon-exchange can explain the observed splitting among spin 0 and spin 1 mesons, and spin 1/2 and spin 3/2 baryons. Lipkin \cite{Lipkin:2007cg} often stressed the pioneer  observations by Sakharov on this subject. The best known references are the paper by De R\'ujula, Georgi and Glashow  \cite{DeRujula:1975ge} and the bag model \cite{DeGrand:1975cf}. See, also, \cite{Jaffe:1999ze} for a review. 
We present here the simplest version, suited for potential models, but the conclusions are somewhat more general.

In short, one can start from a quark--antiquark  or three-quark bound state, and describe the hyperfine splitting by
\begin{equation}
 \label{eq:ss}
V_{\rm ss}=-A\,\sum_{i<j} v_{ss}(r_{ij})\,\frac{\tilde\lambda_i.\tilde\lambda_j\,\vec\sigma_i.\vec\sigma_j}{m_i\,m_j}~,
\end{equation}
where the $m_i$ are the constituent masses, $\vec \sigma_i$ the spin operator, $\tilde\lambda_i$ the color operator, with suitable change for  antiquarks, which belong to the conjugate representation $\bar 3$ of SU(3), and $v_{ss}$ a short-range operator, most often taken as a mere delta function. In the one-gluon-exchange model, the strength $A$ is related to the QCD coupling constant $\alpha_s$.  For a meson, $\langle\tilde\lambda_1.\tilde\lambda_2\rangle=-16/3$, and then the vector to pseudoscalar mass difference is given, at first order, by
\begin{equation}
 \label{eq:dmmeson}
M({}^3{\rm S}_1)-M({}^1{\rm S}_0)=\frac{64\,A}{3}\,\langle v_{\rm ss}\rangle~,
\end{equation}
and the analog for $(qqq)$ baryons is given by
\begin{equation}
 \label{eq:dmbaryon}
M(3/2)-M(1/2)=\frac{48\,A}{3}\,\langle v_{\rm ss}\rangle~,
\end{equation}
where one uses $\langle\vec\sigma_1.\vec\sigma_2\rangle=-3,\,1$ for spin $S=0$ and spin $S=1$, respectively, and 
$\sum\langle\vec\sigma_i.\vec\sigma_j\rangle\pm3$ for spin $3/2$ and $1/2$ baryons. Note also that the color factor is reduced by a factor 1/2 for a quark--quark pair in a baryon as compared to a quark--antiquark in a color singlet. Assuming the same spatial matrix element $\langle v_{\rm ss}\rangle$, this simplistic model gives a ratio 3/4 for $(\Delta-N)/(\rho-\pi)$.
The systematics of ground-state baryons and mesons with various quark content is rather impressive, given the crudeness of this approach. In particular, the mass dependence in \eqref{eq:ss} provided the first explanation  of why $\Sigma$ is heavier than $\Lambda$, as in the latter case the light-quark pair has spin 0, instead of spin 1 in the former case \cite{DeRujula:1975ge,LeYaouanc:1976ne}.

In mesons and baryons, the color factor is frozen, and the calculation of hyperfine splittings  reduces to the simple algebra of the spin--spin operator $\vec\sigma_i.\vec\sigma_j$. However, in 1977, Jaffe realized that the spin--color operator
$\mathcal{O}=\sum \tilde\lambda_i.\tilde\lambda_j\,\vec\sigma_i.\vec\sigma_j$ exhibits dramatic coherences in some multiquark configurations, with some positive expectation values, i.e.,, attractive contributions if $A>0$ and $V_{\rm ss}>0$, larger than the cumulated values in the hadrons constituting the threshold \cite{Jaffe:1976yi}.

Note that this is a rare property. If on considers  the positronium molecule $(e^+,e^+,e^-,e^-)$, the sum of charge 
factors, $\sum q_i\,q_j=-2$, has the same value as in the threshold $(e^+,e^-)$: there is no obvious excess of attraction, and the two atoms have to polarize each other to dynamically produce a small effective attraction.  

In 
\cite{Jaffe:1976yi}, an estimate was done of the $H=(uuddss)$ hexaquark configuration, assuming \textsl{i)} SU(3) symmetry, and 
\textsl{ii)} that for each quark pair, the expectation value $\langle v_{\rm ss}(r_{ij})\rangle$ is the same in $H$ as in ordinary baryons. Another assumption, less explicitly stated, is that one starts, somehow, from an existing $H$ before switching on the chromomagnetic interaction, and that this hexaquark is degenerate with the threshold made of two baryons.

This gave
\begin{equation}\label{eq:H}
 M(H)-2\,M(\Lambda)=(M(N)-M(\Delta))/2\simeq 150\,\text{MeV}~.
\end{equation}
This $H$ became very fashionable and was searched for in more than 20 experiments. 

In 1987, Lipkin and independently the Grenoble group \cite{Gignoux:1987cn,Lipkin:1987sk}, realized that, with the same hypothesis about SU(3) flavor symmetry and the expectation value $\langle v_{\rm ss}(r_{ij})\rangle$ in the light-quark sector,  the heavy pentaquark with one heavy antiquark and four light quarks, $(\bar Q qqqq)$, is bound by the same amount below the $(\bar Qq)+(qqq)$ threshold. In this pentaquark, $qqqq$ is a triplet of SU(3) with spin 0, i.e., a scalar $uuds$, $udds$ or $udss$. The pentaquark was searched for at Fermilab, but the experiment was not conclusive \cite{Aitala:1997ja}.

Needless to say that the striking prediction \eqref{eq:H} and its pentaquark analog were put under scrutiny by the community. Oka, Shimizu and Yazaki, in particular, attempted a genuine 6-body estimate of the $H$ \cite{Oka:1983ku,Oka:2000wj}. See, also, the analyzes in \cite{Rosner:1985yh,Karl:1987cg,Karl:1987uf,Fleck:1989ff}, were the $H$ and the $P$ appear to be much less stable than in the original papers, and even very likely unstable. The main reasons are: SU(3) breaking does not change much the chromomagnetic attraction in the $\Lambda+\Lambda$ threshold of $H$ or in the $\bar D+\Lambda$ threshold of $P$, while the multiquark losses some attraction; the short-range orbital factors $\langle v_{\rm ss}(r_{ij})\rangle$ are much weaker in a dilute multiquark than in a compact baryon. Note, however, that some recent lattice calculation  \cite{Beane:2010hg,Inoue:2010es,Shanahan:2011su} suggests that $H$ could be either weakly bound or just the above its threshold.

Whatever the final result for $H$ and $P$, it remains that the spin--color operator $\mathcal{O}=\sum \tilde\lambda_i.\tilde\lambda_j\,\vec\sigma_i.\vec\sigma_j$ exhibits interesting properties. Thanks to the overall antisymmetrization, the expectation value of $\mathcal{O}$ also depends on flavor, and the more antisymmetric the wave function in flavor, the better the attraction. This is why this chromomagnetic binding involve flavored quarks. Some systematics of $\langle \mathcal{O}\rangle$ can be found in \cite{Hogaasen:1978xs,Aerts:1979hn}.

A variant of the chromomagnetic interaction \eqref{eq:ss} is a spin--flavor interaction, where the operator becomes proportional to $\sum\vec\sigma_i.\vec\sigma_j\,\vec\tau_i.\vec\tau_j$ for SU(2), easily extended to SU(3)$_\text{F}$. This alternative to hyperfine forces was much advertised in the case of baryons \cite{Glozman:1995fu}. The patterns of splittings are rather similar to the ones induced by the spin--color operator, since spin color and flavor are related by the requirement of overall antisymmetry. Of course, SU(3) breaking is more severe here, as instead of a simple substitution of $m_{u,d}$ by $m_{s}$, one replace the exchange of a pion between quarks by the exchange of a kaon, which is of much shorter range.
See, e.g., \cite{Stancu:1998pr} for a survey of the  application of spin--flavor mechanisms to multiquark spectroscopy.
\section{Chromoelectric binding}
Another way of building stable multiquark states relies on the spin-independent interaction, or say, the chromoelectric potential, and its property of flavor independence. The gluons are coupled to the color of each quark. Therefore, this is approximately the same potential that acts on $s$, $c$ or $b$ quarks.
For instance, shortly after the discovery of $\Upsilon$ states, interpreted as $(b\bar b)$, potential models were built, which reproduce simultaneously charmonium $(c\bar c)$ and bottomonium $(b\bar b)$. 

The situation is thus very similar to the physics of exotic atoms and molecules, where the very same Coulomb interaction binds electrons, muons, protons and antiprotons, etc. Hence some  guidance for tetraquarks can be sought from the stability of $(m_1^+,m_2^+,m_3^-,m_3^-)$ in atomic physics, which is reviewed in \cite{ARV}.
For equal masses, the positronium molecule is found to be stable against any dissociation, in particular the decay into two separated positronium atoms, which is the lowest threshold. The same stability holds for any rescaled version, such as $(\mu^+,\mu^+,\mu^-,\mu^-)$. The stability of positronium molecule  was guessed in 1945, demonstrated in 1947 and confirmed in several further studies, but its indirect experimental evidence occurred only in 2007! 

If any of the many symmetries of the  $(m^+,m^+,m^-,m^-)$ molecule is broken, its energy decreases, according to a general rule in quantum mechanics. But this does not mean that its stability is improved, as, in most cases, the threshold energy also decreases, and by a larger amount. If for instance, you breaks the symmetry of particle identity, for simplicity in the same manner in both charge sectors, i.e., consider  $(M^+,m^+,M^-,m^-)$, stability is lost for $M/m\gtrsim 2.2$ (or, of course, $m/M\lesssim2.2$). Indeed, a compact $(M^+,M^-)$ atom cannot be enough polarized by a light $(m^+,m^-)$ to bind the two atoms. 

On the other hand, breaking charge conjugation, i.e., going from the equal-mass case 
$(\mu^+,\mu^+,\linebreak[2]{\mu^-,\mu^-)}$.to the asymmetric $(M^+,M^+,m^-,m^-)$ case by keeping $1/M+1/m=2/\mu$ constant, benefits from a combination of two effects:
\textsl{i)} the energy of the four-charge system decreases under symmetry breaking, 
\textsl{ii)} the threshold energy remains constant (since the reduced mass of the $(M»^+,m^-)$ atoms is fixed).
It is, indeed, well-known that the hydrogen molecule is more deeply bound than the positronium one, with a much richer spectrum of excitations. Similarly, in any flavor independent potential, pairwise or more complicated, it is found that $(Q,Q,\bar q,\bar q)$ becomes stable against dissociation, provide the quark-to-antiquark mass ratio $M/m$ is large enough. For a review on the early papers, and some recent developments, see, e.g., the talk given by Vijande at FB20 the week before this SNP12 workshop \cite{Vijande:2012jw}.

\begin{figure}[!!!ht]
 \centering
 \raisebox{.5cm}{\includegraphics[scale=.9]{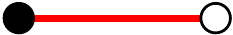}}\qquad
\includegraphics[scale=.9]{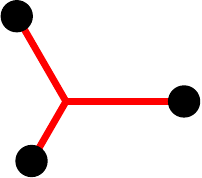}\qquad
\includegraphics[scale=.9]{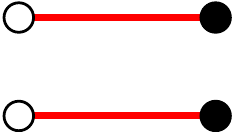}\qquad 
\includegraphics[scale=.9]{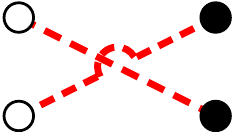}\qquad 
\includegraphics[scale=.9]{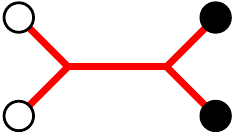}
 \caption{Schematic picture of the string limit of quark--antiquark, three-quark and tetraquark confinement. 
 \label{Fig1}}
\end{figure}

Early studies on tetraquarks were based on empirical 2-body potential $V\propto\sum\tilde\lambda_i.\lambda_j\,v(r_{ij})$, where $v(r)$ is the quarkonium potential. For baryons, this ansatz gives a potential
$[v(r_{12})+v(r_{23})+v(r_{31})]/2$, sometimes referred to as deduced by the ``1/2'' rule form the meson potential. In modern baryon spectroscopy, one restricts this 1/2 rule to the short-range part and describes the linear part by a $Y$-shape interaction: if $\sigma \,r$ is the confining part of the meson potential, the analog for baryons reads $\sigma\,\min_a(r_{1a}+r_{2a}+r_{3a})$ corresponding to three flux tubes joining each quark to a junction $a$ whose location minimizes the potential. See Fig.~\ref{Fig1}.
%

The analog for multiquark confinement includes connected Steiner trees, such as the last diagram in Fig.~\ref{Fig1}, in which the location of all junctions are optimized, and also the so-called ``flip-flop'' terms: the interaction is restricted among clusters, but as the quark move, the content of the clusters is adjusted to minimize the overall potential. For tetraquark, the confining interaction pictured in Fig.~\ref{Fig1} reads
\begin{equation}
 \label{eq:ff4}
V=\sigma\,\min\left[r_{13}+r_{24},\,r_{14}+r_{23},\,\min_{a,b}(r_{1a}+r_{2a}+r_{ab}+r_{b3}+r_{b4})\right]~.
\end{equation}
For both the color-additive model and the more elaborate string-inspired potential, it is observed that $(Q,Q,\bar q,\bar q)$ becomes bound when the mass ratio increases. Remarkably, the same conclusion  is reached in all constituent-model calculations, and is also supported by lattice and sum rules.  It seems thus urgent to investigate more seriously the double-charm sector in experiments, and also the charm-and-beauty and the double-beauty sectors. It is rather puzzling that double-charm baryons, tentatively seen at Fermilab in the Selex experiment \cite{Mattson:2002vu,Ocherashvili:2004hi}, have not been confirmed in other experiments such as BaBar. Meanwhile, double $c\bar c$ production has been seen in $e^+e^-$ collisions, leading to final states such as $J/\psi+\eta_c$. It should be understood why, if two $c\bar c $ pairs are produced, they cannot show up as two $c$ on one side, and two $\bar c$ on the other side. Then final states with a double-charm baryon, an antiproton and two $\bar D$ mesons could be seen. Or, more interestingly, a final state with a $T_{cc}$ tetraquark recoiling against two $\bar D$.

\section{Outlook}
The transition from a naive color-additive model to the string inspired interaction has opened interesting perspectives. Studies have been carried out for the tetraquark \cite{Carlson:1991zt,Vijande:2007ix,Ay:2009zp},  pentaquark \cite{Richard:2009rp} and both $(q^3Q^3)$ and $(Q^3\bar q^3)$ hexaquark  \cite{Vijande:2011im} configurations. However, the treatment of color is rather crude. When the minimum is taken of several string configurations, as per Eq.~\eqref{eq:ff4}, the internal color changes. This is fine to the extent that there is no antisymmetrization constraint and the color degree of freedom can be integrated out in a Born--Oppenheimer approximation. We are presently checking the validity of this approaches, and trying to elaborate an alternative formulation of the string interaction as an operator in color space.

\subsection*{Acknowledgments}
I would like to thank the organizers for the pleasant and stimulating atmosphere  of this workshop, and A.~Valcarce and J.~Vijande for an enjoyable collaboration on the physics issues discussed in this review. 
%

%
\providecommand{\href}[2]{#2}\begingroup\raggedright\endgroup
\end{document}